\documentclass[nofootinbib,prl,amsmath,amssymb,twocolumn,floatfix]{revtex4-2}
\usepackage{multirow}
\usepackage{romannum}
\usepackage{color}
\newcommand{\bfkappa}{\boldsymbol{\kappa}}
\usepackage{graphicx}
\usepackage{dcolumn}
\usepackage{bm}
\usepackage{siunitx}
\usepackage[normalem]{ulem}
\usepackage{xcolor}
\usepackage{soul}
\usepackage{xcolor}

\usepackage{fixme}
\fxsetup{status=draft} 

\begin{document}

\title{Localized Topological States beyond Fano Resonances via Counter-Propagating Wave Mode Conversion in Piezoelectric Microelectromechanical Devices}

\author{Jacopo M. De Ponti$^{1}$, Xuanyi Zhao$^{2}$,  Luca Iorio$^{1}$, Tommaso Maggioli$^{2}$, Marco Colangelo$^{2}$, Benyamin Davaji$^{2}$, Raffaele Ardito$^{1}$, Richard V. Craster$^{3,4}$, Cristian Cassella$^{2*}$}

\affiliation{$^1$ Dept. of Civil and Environmental Engineering, Politecnico di Milano, Piazza Leonardo da Vinci, 32, 20133 Milano, Italy}
\affiliation{$^2$ Electrical and Computer Engineering Department, Northeastern University, Boston, US}
\affiliation{$^3$ Department of Mathematics, Imperial College London, London SW7 2AZ, UK}
\affiliation{$^4$ Department of Mechanical Engineering, Imperial College London, London SW7 2AZ, UK}
\def\andname{} 
\affiliation{Corresponding Author: c.cassella@northeastern.edu}




\begin{abstract}

A variety of scientific fields like proteomics and spintronics have created a new demand for on-chip devices capable of sensing parameters localized within a few tens of micrometers. 
Nano and microelectromechanical systems (NEMS/MEMS) are extensively employed for monitoring parameters that exert uniform forces over hundreds of micrometers or more, such as acceleration, pressure, and magnetic fields. However, they can show significantly degraded sensing performance when targeting more localized parameters, like the mass of a single cell. To address this challenge, we present a new MEMS device that leverages the destructive interference of two topological radiofrequency (RF) counter-propagating wave modes along a piezoelectric Aluminum Scandium Nitride (AlScN) Su-Schrieffer-Heeger (SSH) interface. 
The reported MEMS  device opens up opportunities for further purposes, including achieving more stable frequency sources for communication and timing applications.

\end{abstract}

\maketitle


\noindent \textbf{Introduction.} Emerging needs in proteomics have created a necessity to sense parameters localized within a few tens of micrometers or less. This is crucial for studying biological processes at a single-cell scale \cite{gil-santos_optomechanical_2020, zajiczek_nano-mechanical_2016, maloney_nanomechanical_2014, grover_measuring_2011, andersson_microtechnologies_2004,ahmad_situ_2008,ilic_single_2001}, identifying protein biomarkers associated with specific diseases \cite{hanay_single-protein_2012,yaari_perception-based_2021,mao_highly_2011,cui_nanowire_2001,cai_molecular-imprint_2010}, achieving atomic-scale resolutions in mass spectrometry \cite{sansa_optomechanical_2020,ekinci_ultimate_2004,chaste_nanomechanical_2012,gil-santos_nanomechanical_2010,zobenica_integrated_2017,naik_towards_2009,thakur_real-time_2019}, and more; similar needs have emerged even in other fields of study \cite{quantum_acoustics_2020}. For instance, being able to sense spin waves transduced in strongly localized regions  \cite{spin_current_generation_2020}  can lead to higher bit densities and improved energy efficiency in high frequency spintronic memory devices. Similarly, highly localized acoustic modes of vibration in piezoelectric films can provide a path toward robust quantum state transfer (QST) between remote superconducting qubits  \cite{robust_quantum_state_transfer_2020} and their successful read-out.
Over the past fifteen years, nano and microelectromechanical systems (NEMS/MEMS) have been extensively utilized for sensing parameters (magnetic field \cite{nan_self-biased_2013, mbarek_highly_2021,zaeimbashi_ultra-compact_2021}, acceleration \cite{xu_programmable_2020, kononchuk_exceptional-point-based_2022,middlemiss_measurement_2016,mustafazade_vibrating_2020}, pressure \cite{morten_resonant_1992,welham_laterally_1996,li_ultra-sensitive_2007,venstra_nanomechanical_2014}, etc.) that exert nearly uniform forces across hundreds of micrometers along their frequency-setting dimension. However, these devices are significantly limited in their ability to monitor parameters localized within a few tens of micrometers or less (i.e., in their ability to achieve a high ``spatial sensing resolution").  In fact, to enable high responsivity in such scenarios, the size of NEMS/MEMS must be shrunk to confine their mode of vibration within an area matching closely the one where the targeted parameter exerts its force. However, shrinking the size of the current MEMS/NEMS, particularly along their frequency-setting dimension, results in a notable reduction of their quality factor ($Q$) and dynamic range \cite{ekinci_ultimate_2004}. This reduction amplifies the amount of frequency fluctuations caused by noise \cite{high_figure_of_merit_nems_2020}. Since, the read-out of NEMS/MEMS typically consists of monitoring their resonance frequency, an increase in the amount of noise-driven frequency fluctuations inevitably leads to a degradation in their achievable limit of detection.

Several physical phenomena have been exploited to enhance the spatial sensing resolution of NEMS/MEMS while preserving a high quality factor. These include internal-resonance \cite{xia_internal_2021, wang_frequency_2022, hajjaj_combined_2022}, phase-synchronization \cite{pu_anomalous_2021, xu_fast_2022,defoort_amplitude_2022}, phonon-cavity \cite{miao_nonlinearity-mediated_2022,asadi_nonlinear_2018,okamoto_coherent_2013,zhao_toward_2019}, and mode-localization \cite{spletzer_ultrasensitive_2006,zhao_review_2016,peng_sensitivity_2020}. Most of these phenomena originate from the nonlinear interaction between multiple  mechanical modes. As a result, they necessitate precise control of these modes' driving conditions. Harnessing these phenomena also requires the use of amplitude read-out schemes, which are inherently more susceptible to accuracy degradations caused by electrical noise than the frequency read-out schemes typically used for linear NEMS/MEMS \cite{wang_decouple-decomposition_2023}.

\begin{figure*}[t!]
    \centering
    \includegraphics[width = 1\textwidth]{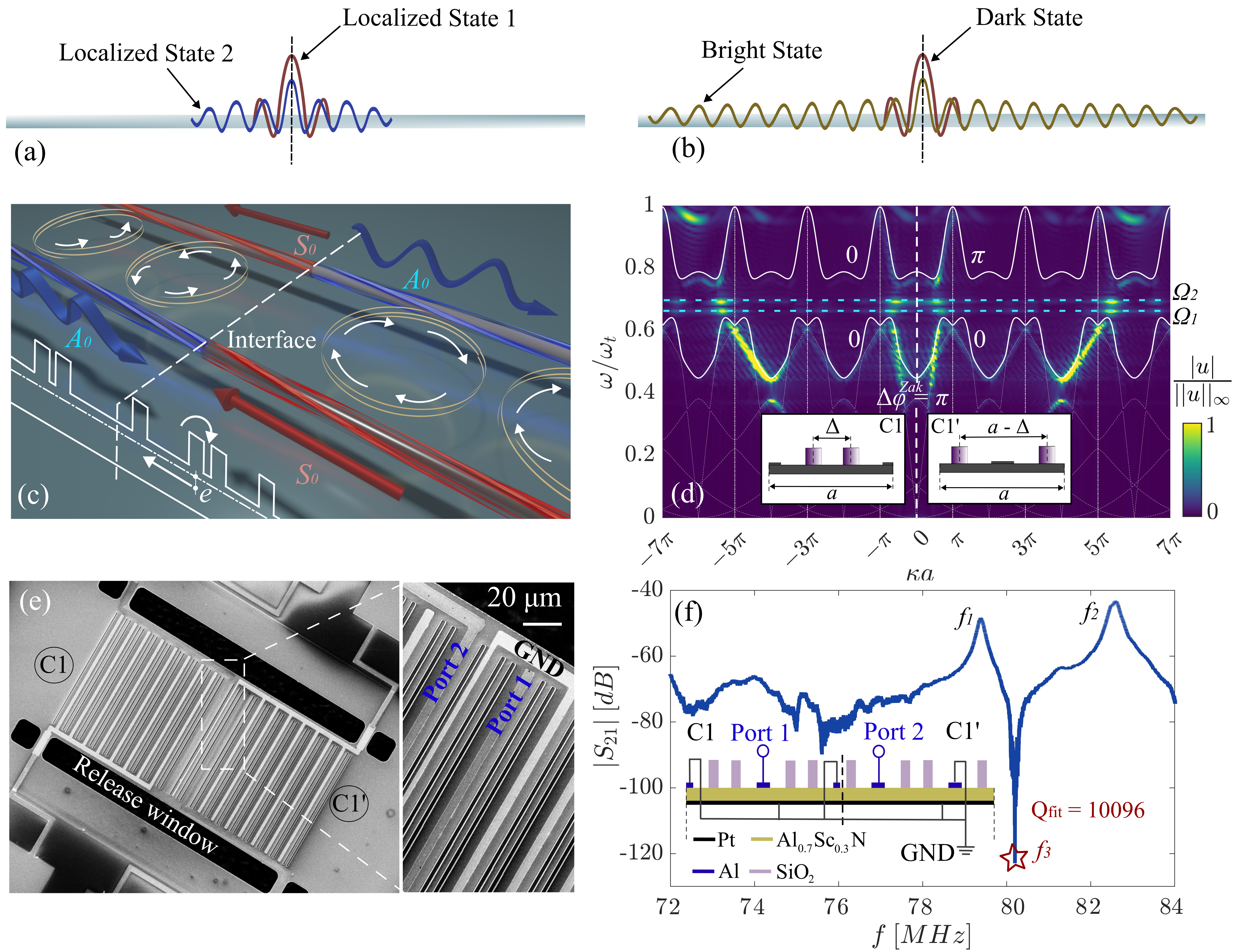}
    \caption{Schematic of the physics and the reported MEMS device, aided by theoretical and experimental results. (a) Strong wave localization is achieved by leveraging the interference of two topologically protected localized states, named as State-1 and State-2. (b) This scenario differs from the conventional Fano resonance, wherein the dark and bright states have very different lifetimes, i.e. the bright state approaches a continuum state and the dark mode is a localized state. (c) The interference between State-1 and State-2  is described by coupling, at a topological interface, counter-propagating symmetric ($S_0$) and antisymmetric ($A_0$) Lamb waves. (d) Such wave coupling is achieved, in one-dimensional wave propagation, by breaking the horizontal spatial symmetry, i.e. by corrugating one side of the surface of an elastic plate. This structure is then combined with an elastic version of the SSH model, where an interface is encountered between two periodic structures (C1 and C1$^\prime$). The unit cells of each structure shares the same periodicity $a$, but have corrugated elements separated by $\Delta$ and $a-\Delta$ respectively. This creates two topological interface states  with distinct Zak phases and different wavenumbers, as shown by the 2D Fast Fourier Transform of wavefield from numerical simulations. (e) We report a piezoelectric MEMS prototype using Aluminium Scandium Nitride as a piezoelectric material, Silicon Oxide to form the periodic corrugations of C1 and C1$^\prime$, and Aluminum strips to form the electrical ports. (f) The experimental electrical transmission of this MEMS device, expressed in terms of the $S_{21}$ scattering parameter, shows the existence of two interface states that lie within a complete bandgap. At their destructive interference, marked with a red star, strong localization is achieved, as well as a record-high $Q$ for MEMS devices using AlScN films as piezoelectric layers.}
    \label{Fig1}
\end{figure*}

In this Article, we demonstrate that both a strong mode-localization and a high quality factor can be simultaneously achieved in a MEMS device operating in a linear regime by leveraging the destructive interference of two topologically protected  states. These states, from now on labeled \textit{State-1} and \textit{State-2} (Fig. \ref{Fig1}-a), 
show comparable lifetimes. Also, they are both localized around an interface, which marks a significant difference with the dark and bright states  (Fig. \ref{Fig1}-b) forming a Fano resonance (i.e. the bright state approaches a continuum state and the dark state is a localized state) \cite{Miroshnichenko2010, Limonov2017}.  

The interference between State-1 and State-2 is described by coupling counter-propagating symmetric $S_0$ and antisymmetric $A_0$ Lamb waves \cite{Lamb1917} at a topological interface in an elastic waveguide (Fig. \ref{Fig1}-c). Such wave coupling is achieved in one-dimensional wave propagation by breaking the horizontal
spatial symmetry, i.e. corrugating one side of the surface of an elastic plate. This system is then combined with an elastic
version of the SSH model \cite{Su1979}, where an interface is encountered between two periodic structures referred to as C1 and C1$^\prime$. The unit cells of each structure shares the same periodicity $a$, but have corrugated elements separated by $\Delta$ and $a-\Delta$, respectively. This creates two topological
interface states with distinct Zak phases and different wavenumbers, as shown by the 2D Fast Fourier Transform of wavefield from
numerical simulations (see Fig. \ref{Fig1}-d).  
We show that the interaction between State-1 and State-2  creates an interference state (State-3) even more localized than both State-1 and State-2. The stronger localization of State-3 enables a significant reduction in radiation losses (also known as anchor losses for NEMS/MEMS devices \cite{nanoscale_imaging_2020}) towards the surrounding silicon substrate. As a result, State-3 exhibits a $Q$ substantially higher than States-1,2. 
It is also important to emphasize that States-1-3 exist within a topologically protected bandgap. This provides immunity to the typical defects \cite{Moore2010, Huber2016} impacting the microfabrication of chip-scale devices. 


To verify the generation of States-1-3, as well as the strong localization and high-$Q$ of State-3, we have built a MEMS device using a piezoelectric Aluminum Scandium Nitride (AlScN) layer (Fig. \ref{Fig1}-e). This device exhibits a quality factor at the resonance frequency of State-3 (Fig. \ref{Fig1}-f) five times superior to any previously reported device made of a thin AlScN film. Moreover State-3 shows a superior localization, with most of its energy being localized in less than 21 $\mu$m along the frequency setting dimension. 

\noindent \textbf{Results} 
The MEMS device demonstrated in this study consists of two periodic structures (i.e., C1 and C1$^\prime$) formed by SiO$_2$ grooves deposited atop a suspended thin bilayer plate. This plate embodies a piezoelectric AlScN layer and a platinum (Pt) layer. C1 and C1$^\prime$ have the same pitch of the unit cell ($a$) but the position of their SiO$_2$ grooves is shifted by an amount equal to $\Delta$ and $a-\Delta$ respectively, as shown in the inset of Fig. 1-d.  This allows us to synthesize an elastic version of the SSH-model. Two 100 nm-thick Al metallic strips are used to piezoelectrically excite the device. These strips, each one forming an electrical port (Port-1 or Port-2), have been positioned near the interface of C1 and C1$^\prime$ to ensure strong enough piezoelectric transduction efficiency to successfully validate the presence of States-1-3 from the extraction of the device's electrical scattering parameters (i.e., the S-parameters). Other grounded Al strips have been inserted across the device to preserve the same dispersion characteristics for all unit cells. Four anchors are used to support the fabricated device after its structural release (i.e., after the removal of the silicon underneath the device to generate a suspension). A Scanned Electron Microscope (SEM) picture of the fabricated device is reported in Fig. \ref{Fig1}-e while the device's electrical transmission (i.e., it $S_{21}$ scattering parameter) is anticipated in Fig. \ref{Fig1}-f. 

\begin{figure*}[t!]
    \centering
    \includegraphics[width = 1\textwidth]{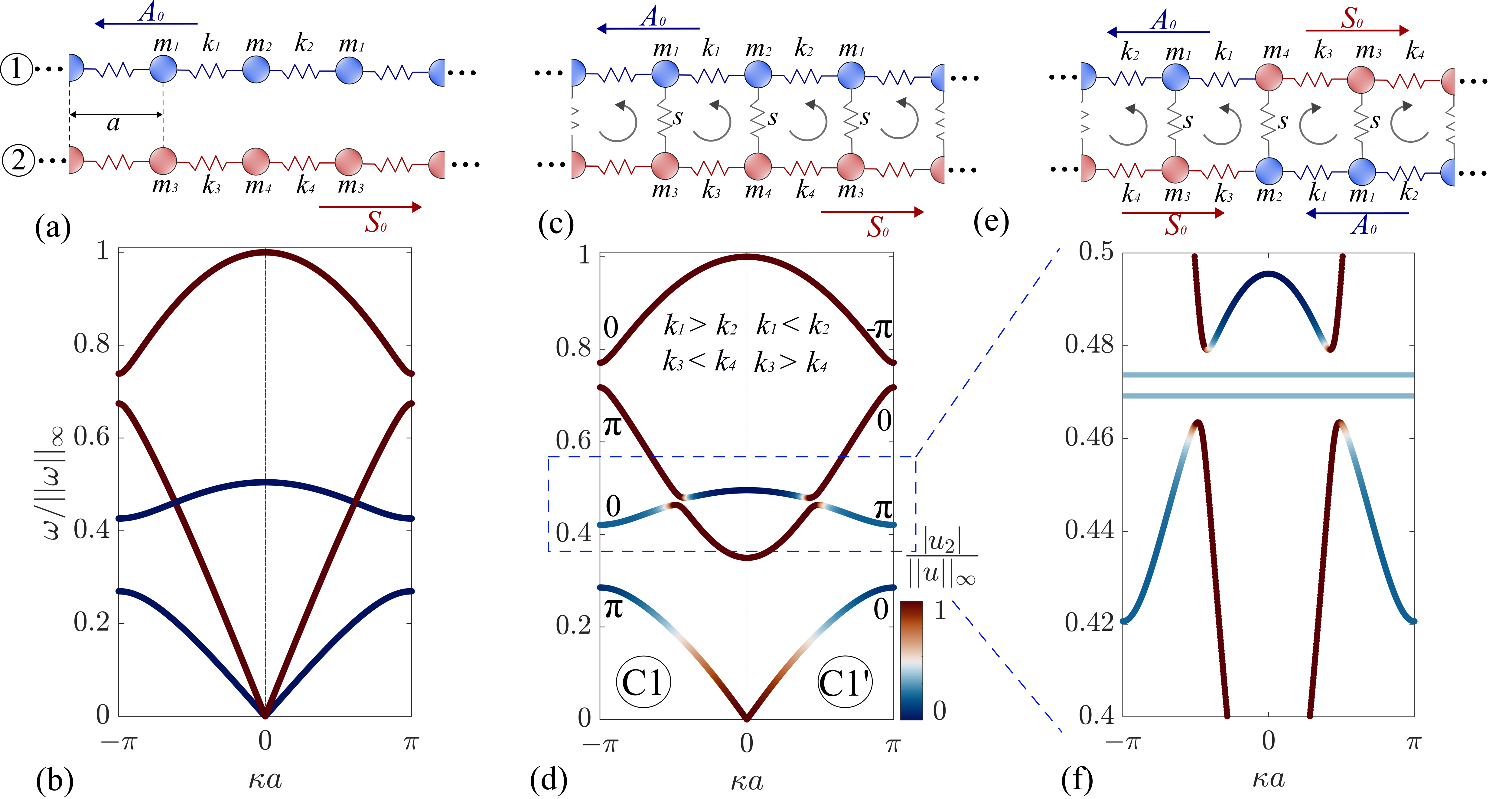}
    \caption{The creation of topological interface states via counter-propagating wave mode conversion is analytically described using a spring-mass model. (a) Chain 1 (blue) models the propagation of antisymmetric $A_0$ waves while chain 2 (red) models the propagation of symmetric $S_0$ waves. (b) When the chains are disconnected, no coupling occurs and band crossing is observed. (c) The introduction of elastic connections couples the two modes, opening a complete bandgap (d) where distinct Zak phases guarantee the existence of non-trivial topological states. (e) The introduction of an interface between the initial chain, having stiffness $k_1$-$k_2$, $k_3$-$k_4$ and its mirrored counterpart, i.e. $k_2$-$k_1$, $k_4$-$k_3$, enables two topological states, here demonstrated from a supercell dispersion analysis (f). The color map of the dispersion curves shows the relative polarization of the waves, with blue points corresponding to vertical (out-of-plane) polarization, while red refers to horizontal (in-plane) polarization, i.e. $A_0$ and $S_0$ waves.}
    \label{Fig2}
\end{figure*}

To explain the origin of States-1,2 in the reported device, we present a toy model which idealizes the propagation of different waves along an elastic substrate. We start by considering two independent periodic spring-mass chains (Fig. \ref{Fig2}-a), each one defined by a set of masses and springs (i.e., $m_1$, $m_2$, $k_1$, $k_2$ and $m_3$, $m_4$, $k_3$, $k_4$ respectively). These chains mimic the propagation of the $A_0$ and $S_0$ wave modes when no coupling between them occurs.  The solution to the wave-equation for the system in Fig. \ref{Fig2}-a can be found by solving the time-independent Schrödinger equation for any possible Bloch state:

\begin{align}
\hat{\boldsymbol{H}}_{\bfkappa}|\boldsymbol{\psi_{{\bfkappa} n}}\rangle = {E}_{{\bfkappa} n}|\boldsymbol{\psi_{{\bfkappa} n}}\rangle.
\label{eq1}
\end{align}

In Eq. (1), ${E}_{{\bfkappa}n}$ identifies a time-independent eigenvalue relative to the wavevector $\bfkappa$ for a specific propagating $n^{th}$ mode. Similarly, 
$|\boldsymbol{\psi_ {{\bfkappa}n}}\rangle$ denotes the solution to the wave-equation for the $n^{th}$ mode and wavevector $\bfkappa$.

The solution of Eq. (1) for the independent chains in Fig. \ref{Fig2}-a is reported in Fig. \ref{Fig2}-b in terms of dispersion curves relating frequency and wavenumber. 
In order to model the coupling between the $A_0$ wave and the $S_0$ wave in our structure, we add connecting springs between the two chains, as shown in Fig. \ref{Fig2}-c.
The resulting dispersion curves are reported in Fig. \ref{Fig2}-d,  where a complete bandgap appears due to the coupling of the two modes. The full $\hat{\boldsymbol{H}}$ matrix, its perturbation and the model parameters for the case study described in  Fig. \ref{Fig2} are reported in the analytical section of the Supplementary Material. 
Next, we can look at the case where an SSH model is reconstructed from the two chains described in Fig. \ref{Fig2}-a,c. To do so, we connect each chain in Fig. \ref{Fig2}-c with a mirrored version forming the C1$^\prime$ structure (Fig. \ref{Fig2}-e). The calculation of the Zak phase  \cite{Zak1989} for both C1 and C1$^\prime$ allows us to verify that non trivial interface states localized at the common interface exist, and these states are topologically protected, being the bandgaps of C1 and C1$^\prime$
endowed with distinct Zak phases. Details about this calculation are reported in the analytical section of the Supplementary Material. The dispersion curves of the supercell, reported in Fig. \ref{Fig2}-f for the system shown in Fig. \ref{Fig2}-e clearly demonstrate the existence of two interface states inside the topological bandgap. As both states have mixed polarizations and store their entire elastic energy within few unit cells around the same interface discontinuity, they can interfere. As a result, an interference state arises, as expected (see Fig. S1) from the analysis of the spring-mass chain system shown in Fig. \ref{Fig2}-e. The modal characteristics of the interference state have been investigated both numerically and experimentally for the MEMS device built in this work, as described in the following. 

\begin{figure*}[t]
    \centering
    \includegraphics[width = 1\textwidth]{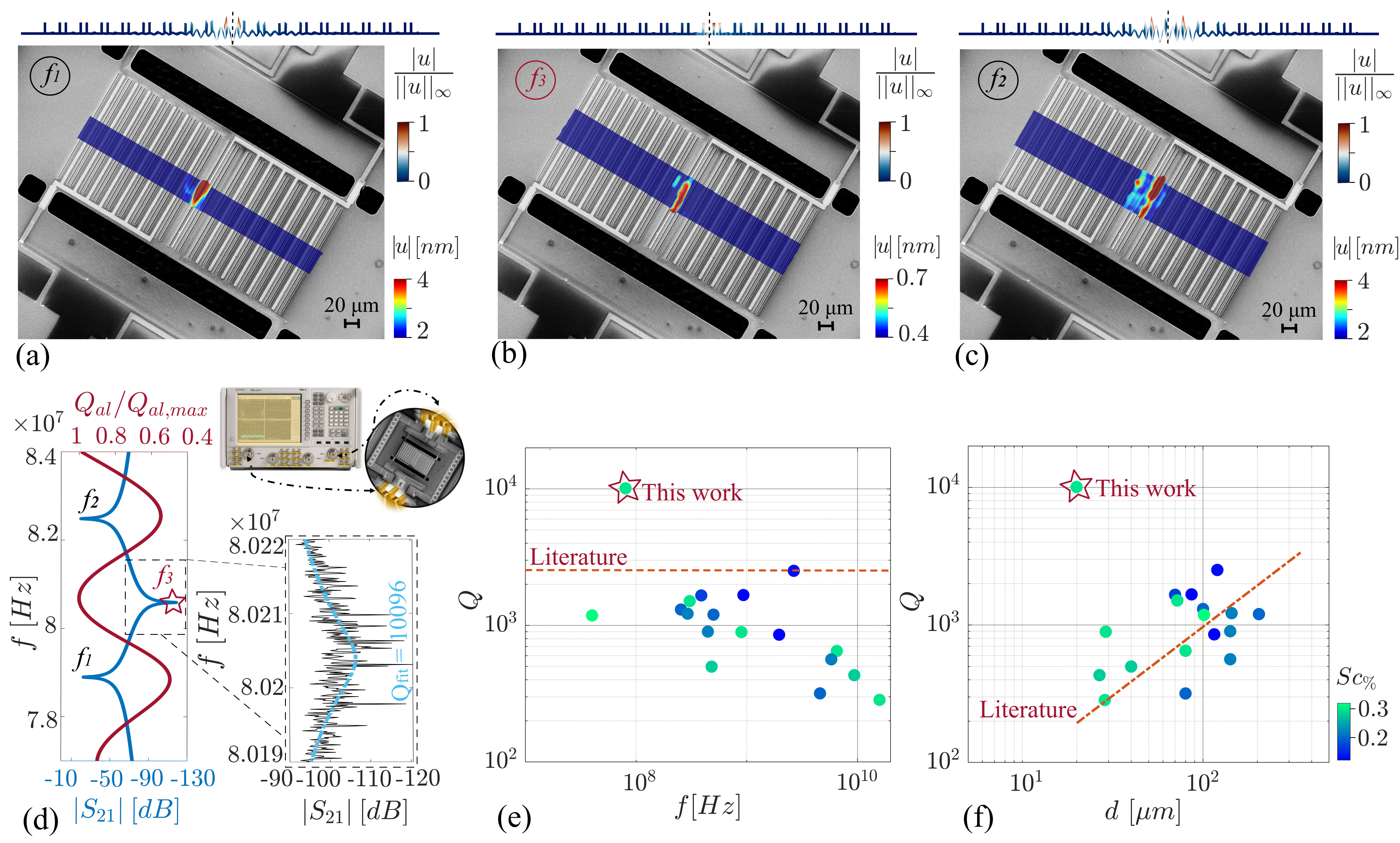}
    \caption{
    (a-c) Measured displacement magnitude for the midsection of the reported MEMS device when operating at $f_1$ (a), $f_2$ (b) and $f_3$ (c). The measurements were performed through a high-frequency vibrometer. Also, corresponding FEM simulated cross-sectional displacement modeshapes are reported above each one of these measured plots. (d) The fabricated MEMS device was electrically tested by probing its input and output ports through two RF GSG probes and by using a Vector Network Analyzer (VNA). We report the FEM calculated frequency distribution of the device transmission ($S_{21}$, in blue), together with a trend of the normalized quality factor term associated to anchor losses (in red) vs. frequency. Also, a zoomed-in picture of the measured $S_{21}$ around $f_3$ is also reported, together with the analytical fitting line we have generated to quantify the MEMS device's $Q$ at $f_3$. (e) A comparison of the $Q$ achieved by the reported MEMS device (see the red star) with those previously reported by state-of-the-art AlScN devices. (f) The same pool of devices has been also used to analyze the $Q$ attained by our reported device vs. the size of its resonant cavity (i.e., $d$) with the corresponding values for each previously reported resonator listed in (e). Evidently, the MEMS device reported in this work not only shows a superior quality factor, but also demonstrates the capability of overruling the current strict correlation of $Q$ with the effective size of the resonant cavity in AlScN MEMS devices.}
    \label{Fig3}
\end{figure*}

     \label{eq:zak}
We characterized the electrical response of the MEMS device demonstrated in this work by measuring its  $S_{21}$ transmission from 60 MHz to 110 MHz using a Vector Network Analyzer. The device's $S_{21}$ trend shows two peaks and one notch in its electrical transmission (i.e., in its $S_{21}$). The two peaks correspond to States-1,2 while the notch corresponds to State-3. From our measured $S_{21}$ vs. frequency trend, we extracted the $Q$ value of State-3. We found the $Q$ of State-3 to be higher than  10,000, as shown in Fig. \ref{Fig1} -f. 
We characterized the localization of States-1-3 by direct measurement of the reported device's displacement through a high-frequency vibrometer (Fig. \ref{Fig3} - a,b,c). We found the measured displacement for State-3 to be localized within an effective cavity width ($d$, calculated along the frequency setting dimension) of 21 $\mu$m. Such a $d$ value was attained by identifying the width of the region around the interface characterized by displacement magnitudes higher than \(\frac{1}{e} \cdot \mu_{\text{max}}\), where \(\mu_{\text{max}}\) is the maximum displacement and \(e\) is the Neper number.
In turn, States-1,2 exhibit a lower $Q$ (approximately 700) and show a more spread modal energy distribution, with State-1 being more localized than State-2. These findings have been confirmed numerically through additional FEM simulations (see in the top insets of Fig. \ref{Fig3}- a-c the FEM simulated cross-sectional modeshapes of the total displacement for States-1-3).

The strong localization of State-3 allows to obtain an exceptionally high-$Q$, being the leakage of elastic energy into the surrounding silicon substrate minimized. To confirm this, we simulated the device's response through finite element methods (FEM), extracting its $S_{21}$ trend at various frequencies (see  Fig. \ref{Fig3}-d) as well as the trend of its $Q$ term ($Q_{al}$) associated to anchor losses. Our FEM simulations further confirm the presence of States-1,3. Also, they show a significant enhancement in the simulated $Q_{al}$ value for frequencies approaching the resonance frequency of State-3 ($f_3$), as shown in Fig. \ref{Fig3}-d.


\noindent \textbf{Discussion} We have demonstrated the first-ever topological microacoustic device using a suspended AlScN thin-film. This device leverages the interaction of two interface states with hybrid polarizations and comparable localization properties. This allows us to generate an interference state simultaneously exhibiting a strong mode-localization and a high quality factor. We compare in Fig. \ref{Fig3}-e the $Q$ achieved at $f_3$ by the built MEMS device with the $Q$ value of other recent AlScN thin-film MEMS devices operating in the RF range. 
 Evidently, the MEMS device demonstrated in this work shows a quality factor exceeding by more than five times anything reported previously for AlScN MEMS devices. Furthermore, we report in Fig. \ref{Fig3}-f the trend of the measured $Q$ value vs. the size of the region where the modal energy is confined along the frequency setting dimension (i.e., $d$) for each one of the AlScN MEMS device reported in Fig. \ref{Fig3}-e. The $d$ values for all the devices listed in Fig. \ref{Fig3}-f have been extracted from reported modeshape distributions. Evidently, a reduction in $Q$ typically occurs for devices showing a lower $d$ value. However, the MEMS device reported in this work overcomes this limitation by exhibiting a  $Q$ value more than one order of magnitude higher than what reported for AlScN MEMS devices with comparable $d$ values. 

The strong localization and high quality factor exhibited by the reported MEMS device strongly motivate its future use for any on-chip sensing applications demanding high spatial resolutions, high sensitivity and low limit of detection. Furthermore, the high-Q achieved by the reported MEMS device at $f_3$ also creates a new path to achieve ultra-stable frequency synthesizers for timing and frequency conversion, which can be manufactured with conventional semiconductor processes.  

\noindent\textbf{{Methods}} 
The fabrication process of the reported MEMS device starts with the deposition of an AlN/Pt/AlScN stack, which is implemented by reactive co-sputtering in the same multi-target chamber, without breaking the vacuum. Then, we etch the AlN/Pt/AlScN stack to form ``release windows". This step is key as it gives direct access to silicon, which is required to be able to etch the silicon under the device through a XeF2 isotropic etching step (ensuring its suspension). Wet-etching of AlScN is then processed to create vias, allowing to electrically ground the bottom Pt layer. 
Then we sputter and pattern a 2 $\mu$m-thick $SiO_2$ layer via Plasma-enhanced chemical vapor deposition (PECVD) to form the surface corrugations, which have the longest dimension along the out-of-plane direction (i.e., along the anchors' direction). Etching of the $SiO_2$ layer to form the corrugations is implemented through a reactive-ion etching (RIE) step. The Al strips are attained by sputtering and patterning a 150 nm-thick aluminum (Al) layer. 
Finally, we deposit a 300 nm-thick gold (Au) layer through evaporation to cover the vias, routing lines and probing pads, ensuring lower ohmic losses. A process flow chart is provided in the supplementary material, together with the definition of the experimental setup.

\noindent\textbf{{Data availability}} 

\noindent The data that supports the findings of this study are available from the corresponding author upon reasonable request.

\noindent\textbf{{Code availability}} 

\noindent The software ABAQUS\textsuperscript{\textregistered} and COMSOL\textsuperscript{\textregistered} has been used for numerical simulations. Meshing and configuration files for the simulations presented here can be obtained upon reasonable request from the corresponding author.

\noindent\textbf{{Acknowledgements.}}

\noindent J.M.D.P, L.I., R.A. and R.V.C. acknowledge the financial support of the European Union H2020 Future and Emerging Technologies (FET) Proactive Metamaterial Enabled Vibration Energy Harvesting (MetaVEH) project under Grant Agreement No. 952039 in supporting the research activity and the prototype realization. X.Z., T.M., and C.C. were funded by the National Science Foundation (NSF) under grant No. 2034948. Authors acknowledge Prof. Amit Lal and Rohan Sanghvi from Cornell University for access to the Laser Doppler Vibrometer and measurements.

\noindent\textbf{Author Contributions.}
J.M.D.P. and X. Z. contributed equally to the paper; C.C., R.C., J.M.D.P and X. Z. developed the research plan; J.M.D.P.and L.I. conceived the initial design and developed numerical and analytical models; X.Z. designed, modeled through FEM, and fabricated the device;  J.M.D.P., L.I. and R.C developed the theory and the physical interpretation; X.Z., T.M., B. D, and C.C. contributed to the design of experiments; All the authors contributed in writing the paper.

\noindent\textbf{Competing Interests.}

\noindent The authors declare no competing interests.

\bibliographystyle{naturemag}

\end{document}


\title{Localized Topological States beyond Fano Resonances via Counter-Propagating Wave Mode Conversion in Piezoelectric Microelectromechanical devices:\\ Supplementary Information}

\author{Jacopo M. De Ponti$^{1*}$, Xuanyi Zhao$^{2*}$,  Luca Iorio$^{1}$, Tommaso Maggioli$^{2}$, Marco Colangelo$^{2}$, Benyamin Davaji$^{2}$, Raffaele Ardito$^{1}$, Richard V. Craster$^{3,4}$, Cristian Cassella$^{2*}$}

\affiliation{$^1$ Dept. of Civil and Environmental Engineering, Politecnico di Milano, Piazza Leonardo da Vinci, 32, 20133 Milano, Italy}
\affiliation{$^2$ Electrical and Computer Engineering Department, Northeastern University, Boston, US}
\affiliation{$^3$ Department of Mathematics, Imperial College London, London SW7 2AZ, UK}
\affiliation{$^4$ Department of Mechanical Engineering, Imperial College London, London SW7 2AZ, UK}
\def\andname{} 
\affiliation{$^*$ Corresponding Authors: jacopomaria.deponti@polimi.it, zhao.xuan@northeastern.edu}

\maketitle
\beginsupplement

To aid insight into the underlying physics of the localized topological states via counter-propagating wave mode conversion, we provide a detailed description of the analytical and numerical models we have developed to investigate the theory presented in the main manuscript.

\section{Analytical model}

An analytical toy model has been devised to simplify the complex real structure and its behaviour in wave mechanics. The simplest model that can fully describe counter-propagating wave mode conversion, usually named as wave locking \cite{mace2012wave}, is a double bi-atomic chain where the mode of first chain is forced to interact with the mode of the second chain through a periodic coupling which is given, in the simplest case, by elastic springs.

As reported in the main text, the two infinite chains are analytically defined by imposing Bloch-Floquet boundary conditions to the bi-atomic system. The non-perturbed Hamiltonian for the single bi-atomic chain is:

\begin{align}
\hat{\boldsymbol{H}}_{\bfkappa} =\begin{bmatrix}
    (k_1+k_2)/m_1 & -(k_1+k_2 e^{i\bfkappa})/m_1\\
    -(k_1+k_2 e^{-i\bfkappa})/m_2 & (k_1+k_2)/m_2
\end{bmatrix}
\label{eq2}
\end{align}

where $m_1$ and $m_2$ are the particle masses, $k_1$ and $k_2$ the stiffnesses and $\bfkappa$ the wavenumber that goes from $-\pi/a$ to $\pi/a$, being $a$ the lattice constant. 
In order to define the solution for two non-interacting chains, the total Hamiltonian of the system reads:

\begin{widetext}
\begin{equation}
\begin{aligned}
\hat{\boldsymbol{H}}_{\bfkappa} &= 
\begin{bmatrix}
(k_1 + k_2)/m_1 & -(k_1 + k_2 e^{i\bfkappa})/m_1 & 0 & 0\\
-(k_1 + k_2 e^{-i\bfkappa})/m_2 &  (k_1 + k_2)/m_2 & 0 & 0\\
0 & 0 & (k_3 + k_4)/m_3 & -(k_3 + k_4 e^{i\bfkappa})/m_3 \\
0 & 0 & -(k_3 + k_4 e^{-i\bfkappa})/m_4 & (k_3 + k_4)/m_4
\end{bmatrix}
\label{eq:02}
\end{aligned}
\end{equation}
\label{eq3}
\end{widetext}

By introducing this Hamiltonian into the time-independent Schrödinger equation (eq. (1) in the main manuscript), and imposing  Bloch-Floquet boundary conditions, we obtain the dispersion relation in Fig. 2-b of the main manuscript, where band crossing is observed. 
To introduce wave coupling between the two chains, a perturbation of the original Hamiltonian $\hat{\boldsymbol{H}}_{\bfkappa}$ is defined as $\hat{\boldsymbol{H}}^{tot}_{\bfkappa} = \hat{\boldsymbol{H}}_{\bfkappa} + \hat{\boldsymbol{H'}}$, with $\hat{\boldsymbol{H'}}$  equal to:

\begin{equation}
\begin{aligned}
\hat{\boldsymbol{H'}} &= 
\begin{bmatrix}
     s/m_1 & 0 & -s/m_1 & 0\\
     0 & s/m_2 & 0 & -s/m_2 \\
     -s/m_3 & 0 & s/m_3 & 0 \\
     0 & -s/m_4 & 0 & s/m_4
\end{bmatrix},
\label{eq4}
\end{aligned}
\end{equation}

This solution considers now a coupling spring attached to each mass of the bi-atomic chain, allowing for energy flow from one chain to the other. The corresponding dispersion curves are reported in Fig. 2-d of the main manuscript, where a complete bandgap appears due to the coupling of the two states. Table \ref{tab:01} 
reports the parameters used for the analytical results reported in the main manuscript. 


\begin{table}[h]
\centering
\begin{tabular}{|c|c|}
\hline
\textbf{Variable} & \textbf{Value} \\
\hline
$m_1$, $m_2$, $m_3$, $m_4$  & 1, 1, 5, 5 \\
$k_1$, $k_2$, $k_3$, $k_4$   & 1, 1.2, 0.8, 2  \\
s & 0.5  \\
a & 1  \\
\hline
\end{tabular}
\caption{Parameters used in the analytical spring-mass model of Fig.3.}
\label{tab:01}
\end{table}
To investigate the topological features of this bandgap, we  then calculate the Zak phase \cite{Zak1989} for two chains, named as C1 and C1', having stiffness $k_1$-$k_2$, $k_3$-$k_4$ and its mirrored counterpart $k_2$-$k_1$, $k_4$-$k_3$, repectively. This calculation, allows us to verify that, if an interface is encountered between cells C1 and C1', non trivial interface states localized at the common interface exist, and these states are topologically protected, being the bandgaps of C1 and C1' endowed with distinct Zak phases. 
The Zak phase, $\varphi_{n}^{Zak}$, for the $n^{th}$ band is defined in terms of the Berry connection $\mathcal{\chi(\boldsymbol{\kappa})}$ such that

\begin{align}
     \varphi_{n}^{Zak} &= \int_{-\pi/a}^{\pi/a}\chi(\bfkappa)d\bfkappa = i\int_{-\pi/a}^{\pi/a}\langle \psi_n(\bfkappa)|\partial_{\bfkappa}\psi_n(\bfkappa)\rangle,
     \label{eq:zak}
\end{align}

where $a$ is the lattice constant,  $\partial_{\bfkappa}$ is the partial derivative with respect to the wavevector $\bfkappa$ and $\psi(\bfkappa)$ is the eigensolution for the $n^{th}$ band. Since a straightforward implementation of perturbation theory cannot provide the Zak phase as a loop integral of the connection, we implement the more general discrete approach without assuming any regularity of the phase along the path \cite{Resta2000}:

 \begin{align}
 \begin{split}
    \varphi_{n}^{Zak} &= -\mathfrak{Im}\left(\log\prod\limits_{s=1}^{N}\langle \psi_n(\bfkappa_{s})|\psi_n(\bfkappa_{s+1})\rangle\right).
 \end{split}
 \label{eq:zak2}
 \end{align}
 
In Eq. (\ref{eq:zak2}) the periodic gauge condition is satisfied through Bloch condition $|\psi_n(\bfkappa_{N+1})\rangle = e^{-i \boldsymbol{G}\cdotp \boldsymbol{r}}|\psi_n(\bfkappa_{1})\rangle$, being $\boldsymbol{G}$ and $\boldsymbol{r}$ the reciprocal and Bravais lattice vectors respectively.  

Due to the intrinsic connection with Wannier charge centers \cite{Kohn1959,Kivelson1982}, provided that we have inversion symmetry with respect to the array axis, the Zak phase can only assume two values: $0$ or $\pi$. We find distinct Zak phases for all the branches of C1 and C1$^\prime$. As such, we have an analogue of an incomplete Wannier state at the interface between C1 and C1$^\prime$. This state supports topological interface states. The values of the Zak phases for each dispersion branch are reported Fig. 2-d of the main manuscript.

The existence of the two states is also corroborated performing a supercell dispersion analysis, i.e. considering a long chain having an interface between C1 and C1'; the two interface states, with mixed out-of-plane and in-plane polarization, are shown in Fig. 2-d of the main manuscript. We can also numerically investigate the origin of State-3. In this regard, the interference between State-1 and State-2 is analytically demonstrated by analyzing in frequency domain two coupled chains with an interface, as shown in Fig. S1-a. Computing the transmission $\tau$ (Fig. S1-b) between two degrees of freedom close to the interface allows to identify two states at the normalized frequencies $\Omega_1$ and $\Omega_2$ that correspond to States-1,2 of our MEMS device. $\tau$ also shows the presence of an interference state at a normalized frequency $\Omega_3$ that corresponds to State-3 in our MEMS device. Finally, our numerical investigation further demonstrates the stronger localization that the system in Fig. S1-a shows at $\Omega_3$ compared to both $\Omega_1$ and $\Omega_2$ ((Fig. S1-c)).

\begin{figure}[h!]
    \centering
    \includegraphics[width = 1\linewidth]{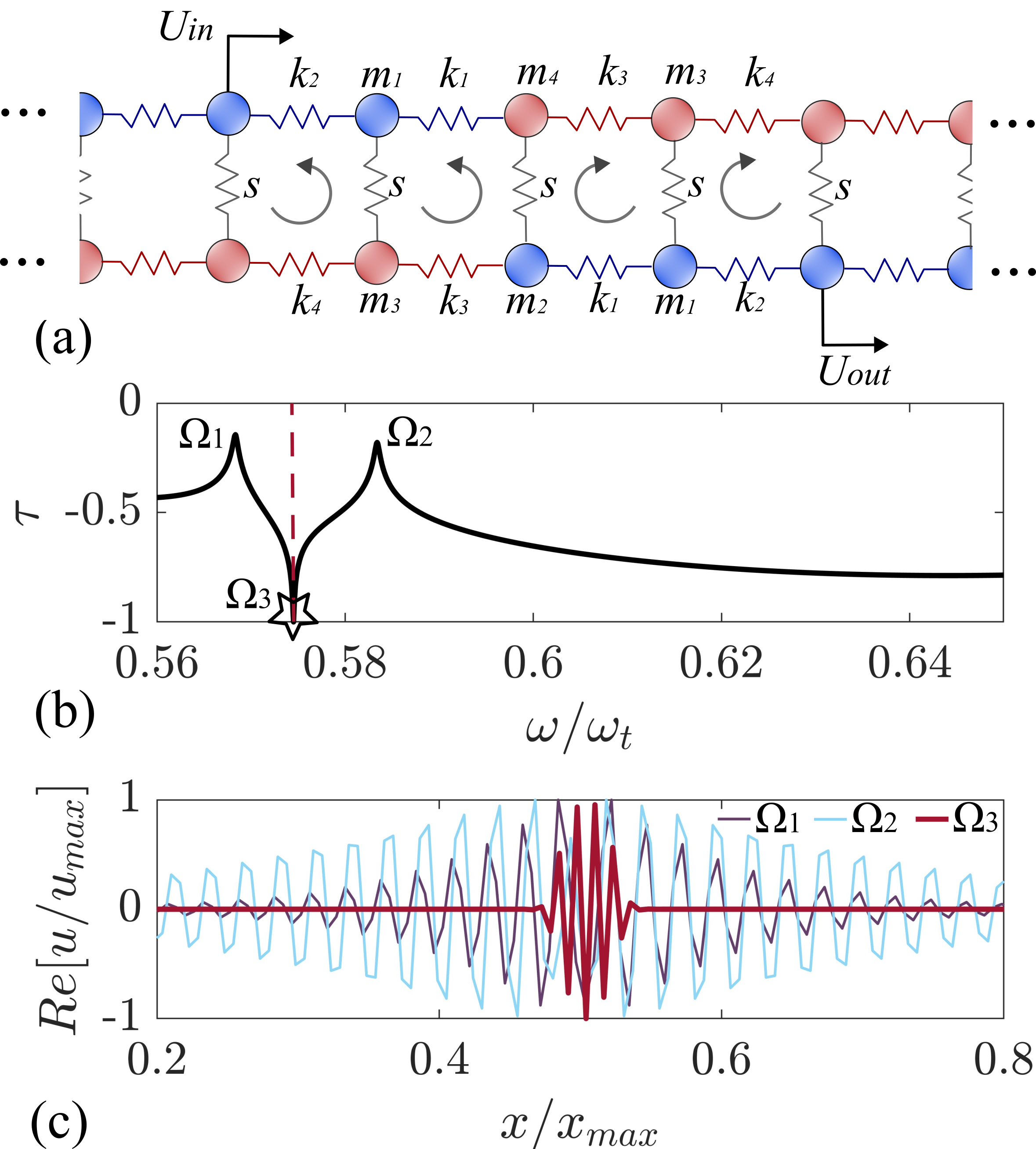}
    \caption{(a) Two coupled spring mass chains with an interface are used to predict the two interface states and their destructive interference. (b) Analytical transmission computed as the ratio between the displacement of two degrees of freedom close to the interference, named as $U_{in}$ and $U_{out}$. (c) Stronger wave localization is observed at $\Omega_3$ compared to both $\Omega_1$ and $\Omega_2$}
    \label{FigS1}
\end{figure}

\section{Numerical analyses}

Numerical analysis are performed using both ABAQUS\textsuperscript{\textregistered} and COMSOL\textsuperscript{\textregistered} FEM software.

To corroborate the existence of the two interface states, we complement the numerical dispersion analyses with a two-dimensional (2D) spatiotemporal Fast Fourier Transform (FFT) of the wavefield obtained from our time domain analyses performed in ABAQUS\textsuperscript{\textregistered}. We adopt a Finite Element discretisation based on 8-node biquadratic plane strain quadrilateral elements (CPE8), with 2 degrees of freedom (dof) for each node. The analysis is performed via implicit time integration based on the Hilber-Hughes-Taylor operator \cite{Implicit}, an extension of the Newmark $\beta$-method with a constant time increment dt = $1$ $ns$. We consider a system with an SSH interface, 10 cells on each side and external absorbing boundary conditions using the ALID (Absorbing Layers using Increasing Damping) method \cite{Rajagopal12}.
The system is excited through an imposed vertical force at the interface, using a linear finite chirp from $40$ to $120$ $MHz$ with a tapered cosine window and time duration of $1$ $\mu s$.
The wavefield, measured in a spatial domain close to the interface, is then 2D Fast Fourier Transformed in order to get the spectrogram reported in Fig. 1(d) of the main manuscript.



    \label{FS_Fab}